\documentclass[aps, prc, reprint, showkeys,superscriptaddress, nofootinbib, floatfix]{revtex4-2}

\usepackage{amsmath}
\usepackage{graphicx}
\usepackage{dcolumn}
\usepackage{bm}
\usepackage{mathtools}

\usepackage[bookmarksnumbered, pdfpagelabels=true, plainpages=false, colorlinks=true, linkcolor=blue, citecolor=blue, urlcolor=blue]{hyperref}

\usepackage{tikz}
\usetikzlibrary{shapes.arrows}
\usetikzlibrary{positioning}
\usetikzlibrary{external}
\tikzexternaldisable
\usepackage{amssymb}
\usepackage{dsfont}
\usepackage{url}
\usepackage{physics}
\usepackage{xcolor}

\usepackage[normalem]{ulem}

\newcommand{\appropto}{\mathrel{\vcenter{
  \offinterlineskip\halign{\hfil$##$\cr
    \propto\cr\noalign{\kern2pt}\sim\cr\noalign{\kern-2pt}}}}}

\begin{document}

\title{Feasibility of measuring the speed of sound of the quark-gluon plasma from the multiplicity and mean $p_T$ of ultracentral heavy-ion collisions}%

\author{Lorenzo Gavassino}
\email{lorenzo.gavassino@vanderbilt.edu}
\affiliation{Department of Mathematics, Vanderbilt University, Nashville, TN 37240, USA}

\author{Henry Hirvonen}
\email{henry.v.hirvonen@vanderbilt.edu}
\affiliation{Department of Mathematics, Vanderbilt University, Nashville, TN 37240, USA}
\affiliation{Department of Physics and Astronomy, Vanderbilt University, Nashville, TN 37240, USA}

\author{Jean-François Paquet}
\email{jean-francois.paquet@vanderbilt.edu}
\affiliation{Department of Physics and Astronomy, Vanderbilt University, Nashville, TN 37240, USA}
\affiliation{Department of Mathematics, Vanderbilt University, Nashville, TN 37240, USA}

\author{Mayank Singh}
\email{mayank.singh@vanderbilt.edu}
\affiliation{Department of Physics and Astronomy, Vanderbilt University, Nashville, TN 37240, USA}

\author{Gabriel Soares Rocha}
\email{gabriel.soares.rocha@vanderbilt.edu}
\affiliation{Instituto de F\'{\i}sica, Universidade Federal Fluminense, Niter\'{o}i, Rio de Janeiro, 24210-346,
Brazil}
\affiliation{Department of Physics and Astronomy, Vanderbilt University, Nashville, TN 37240, USA}

\begin{abstract}
 The mean transverse momentum $\langle p_T \rangle$ of hadrons has been observed experimentally and in numerical simulations to have a power-law dependence on the hadronic multiplicity $N$ in ultracentral relativistic heavy-ion collisions: \mbox{$\langle p_{T} \rangle \propto N^{b_{\rm UC}}$}. It has been put forward that this exponent $b_{\rm UC}$ is the speed of sound of quark-gluon plasma measured at a temperature determined from $\langle p_T \rangle$. We study step by step the connection between (i) the energy and entropy of hydrodynamic simulations and (ii) experimentally measurable observables. We show that an argument based on energy and entropy should yield an exponent equal to the pressure over energy density $P/\varepsilon$, rather than the speed of sound $c_s^2$; however, we also observe that $\langle p_T \rangle$ and $N$ are not sufficiently accurate proxies for the energy and entropy to make this possible in practice. 
 From simulations, we find that the exponent $b_{\rm UC}$ is significantly different whether the ``effective volume'' is strictly constant or not, a condition that cannot be enforced experimentally.
 Additional tests using a modified equation of state find that the exponent $b_{\rm UC}$ exhibits a variable degree of correlations with the speed of sound and with $P/\varepsilon$, but is not an accurate measurement of either quantity in general. 
\end{abstract}

\maketitle

\section{Introduction}
Ultrarelativistic heavy-ion collision experiments at the Large Hadron Collider (LHC) access the high-temperature, low-baryon chemical potential regime of the quantum chromodynamics (QCD) phase diagram \cite{Wong:1995jf,Yagi:2005yb,Vogt:2007zz,Heinz:2013th,Heinz:2024jwu}. Thermodynamic properties of bulk QCD matter produced in these collisions are related to each other by the equation of state. The QCD equation of state has been computed from first principles using lattice QCD techniques \cite{Borsanyi:2010cj,Bazavov:2009zn,Borsanyi:2021sxv,Bazavov:2017dus}. It has also been studied from measurements using Bayesian inference~\cite{Pratt:2015zsa}.

Recently, it was claimed that the QCD equation of state can be directly inferred from experiments in ultracentral collisions \cite{Gardim:2019xjs,Gardim:2019brr,Gardim:2024zvi}. More specifically, the speed of sound squared ($c_s^2$) at an effective temperature, $T_{\rm eff}$, was equated to the experimental observable 
\begin{equation}
b_{\rm UC} \equiv \frac{{\rm d}\ln\langle p_{T,{\rm ch}}\rangle}{{\rm d}\ln N_{\rm ch}},    
\end{equation}
for ultracentral events. Here, $\langle p_{T,{\rm ch}}\rangle$ is the mean transverse momentum of the particles and $N_{\rm ch}$ is their multiplicity. This observable has been measured by the CMS \cite{CMS:2024sgx}, ATLAS \cite{ATLAS:2024jvf} and ALICE \cite{ALICE:2025rtg} collaborations and compared to lattice QCD calculations \cite{HotQCD:2014kol}. 
Studying ultracentral Pb-Pb collisions at $\sqrt{s_{NN}}=5.02$~TeV, the CMS Collaboration found an excellent agreement of $b_{\rm UC}$ with the lattice result for the speed of sound; on the other hand, the ALICE Collaboration observed a large dependence of $b_{\rm UC}$ on the method used for centrality determination. The latter effects were also studied in numerical simulations~\cite{Nijs:2023bzv}.

The purpose of this work is to study the validity and generality of the claim $b_{\rm UC} \overset{?}{=} c_s^2$ in various situations. This paper is organized in the following way. In Sec.~\ref{sec:theory_argument} we go through the original argument and discuss the theoretical difficulties in connecting $b_{\rm UC}$ with the thermodynamic properties of the plasma. The framework that we use to systematically study the observable $b_{\rm UC}$ is introduced in Sec.~\ref{sec:model}. The main results of our work are presented in Sec.~\ref{sec:results}, and final conclusions of our work are given in Sec.~\ref{sec:discussion}.

\section{Connecting $b_{\rm UC}$ to QCD thermodynamics}
\label{sec:theory_argument}

In thermodynamics, the speed of sound squared at zero baryon density is defined as $c_s^2 \equiv {\rm d} P/{\rm d}\varepsilon = {\rm d}\ln T/{\rm d}\ln s$, where $P$ is pressure, $\varepsilon$ is the energy density, $T$ is temperature and $s$ is entropy density. The quark-gluon plasma is not in global thermodynamic equilibrium, but one can define effective thermodynamic variables --- an effective temperature $T_{\rm eff}$ and an effective volume $V_{\rm eff}$  --- through the equalities
\begin{equation}
\label{eq:E_epsilon_Veff}
\begin{split}
    E &\equiv \varepsilon(T_{\rm eff})V_{\rm eff},\\
    S &\equiv s(T_{\rm eff})V_{\rm eff},
\end{split}    
\end{equation}
where $E$ and $S$ are the total energy and total entropy on a constant-temperature hypersurface that encloses a certain spacetime region of the plasma. Because this hypersurface is generally taken to enclose a finite region in spatial rapidity, $E$ and $S$ represent only part of the energy and entropy of the plasma. 
In consequence, the hypersurface energy 
$E$ and entropy $S$ calculated with a such finite rapidity window are \emph{not} conserved quantities --- energy and entropy flow in the longitudinal direction.
Nevertheless, $E$ and $S$ are well defined in numerical simulations. In reality, there are various sources of entropy production in heavy-ion collisions, hence entropy would not be strictly conserved in general even if the entire hypersurface was considered.\footnote{In a previous work~\cite{SoaresRocha:2024drz}, we found that a momentum rapidity coverage of approximately $-10<y<10$ would be needed to recover the full energy and entropy of the plasma, which is beyond the capabilities of existing detectors. We thus emphasize that the non-conserved $E$ and $S$ calculated in a finite rapidity window are the relevant quantities to discuss.}

Two key assumptions are that (i) the medium is in local thermal equilibrium on this hypersurface, and that (ii) the energy $E$ and entropy $S$ on this hypersurface are closely related to the energy and multiplicity of particles measured at the end of the collision. 

\subsection{Speed of sound vs. $P/\varepsilon$}

The original argument for the equivalence between $b_{\rm UC}$ and $c_s^2$ made in Refs.~\cite{Gardim:2019xjs,Gardim:2019brr,Gardim:2024zvi} was the following. 
Guided by the ideal gas example and observations from numerical simulations, Refs.~\cite{Gardim:2019xjs,Gardim:2019brr} state that  $\langle p_T\rangle$ is approximately proportional to the effective temperature $T_{\rm eff}$ \cite{VanHove:1982vk} defined in Eq.~\eqref{eq:E_epsilon_Veff}. Moreover, the particle multiplicity is known to be approximately proportional to the total entropy in the system, so that $N_{\rm ch} \appropto S = s(T_{\rm eff})V_{\rm eff}$. If, moreover, the effective volume $V_{\rm eff}$ is a constant, then $N_{\rm ch}  \appropto s(T_{\rm eff})$ and one finds
\begin{equation}
\label{eq:buc_cs2}
 b_{\rm UC} \equiv   \frac{{\rm d}\ln\langle p_T\rangle}{{\rm d}\ln N_{\rm ch}} 
 \underbrace{=}_{\substack{\langle p_T\rangle  \propto T_{\rm eff}\\ \\ {\rm d}V_{\rm eff}=0 \\ \\ N_{\rm ch} \propto S}}
 \frac{{\rm d}\ln T_{\rm eff}}{{\rm d}\ln s(T_{\rm eff})} = c_s^2(T_{\rm eff})
\end{equation}
In summary, while the last equality is just a thermodynamic relation, the central equality is only valid if the following assumptions hold
\begin{itemize}
    \item[a)] The charged hadron multiplicity $N_{\rm ch}$ is proportional to the hypersurface entropy $S$ ($N_{\rm ch} \propto S$);
    \item[b)] $\langle p_T\rangle  \propto T_{\rm eff}$;
    \item[c)] The effective volume $V_{\rm eff}$ is constant across different ultracentral events. 
\end{itemize}

\begin{figure}
    \centering
    \includegraphics[width=\columnwidth]{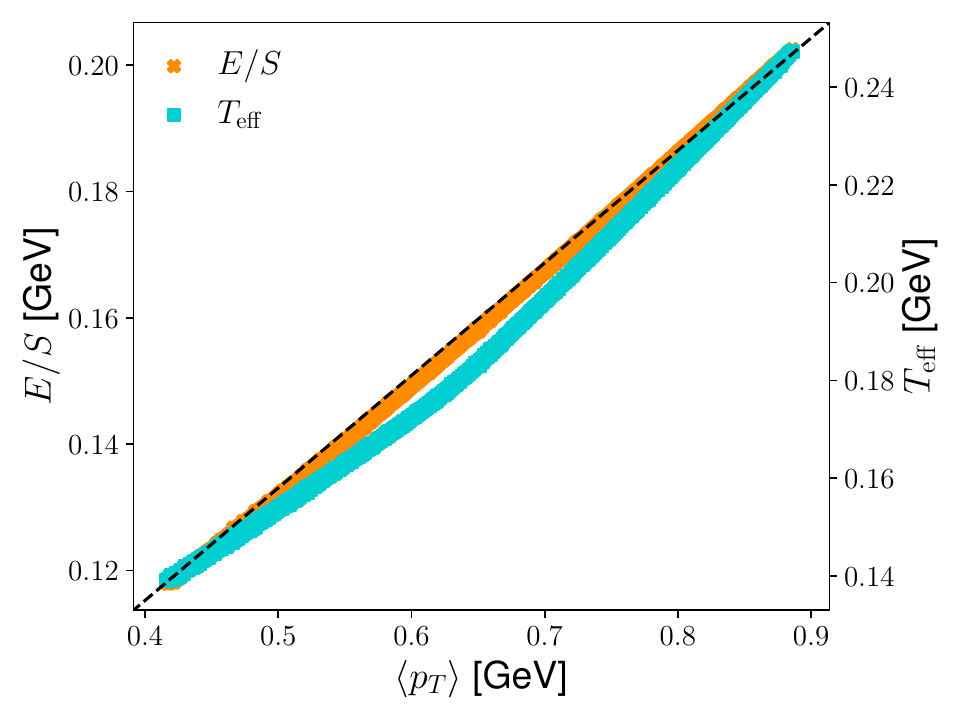}
    \caption{The energy over entropy ratio $E/S$ and the effective temperature $T_{\rm eff}$ computed on a 130 MeV constant temperature hypersurface, as a function of mean transverse momentum of charged hadrons for a QCD equation of state. The dashed black line is a straight line added as a reference.}
    \label{fig:pt_corr}
\end{figure}

Moving beyond the original argument, we point out that a slightly different argument will yield a different interpretation for  $b_{\rm UC}$.
There is no simple analytic expression that connects $\langle p_T\rangle$ with the hypersurface quantities, even in highly symmetric systems~\cite{SoaresRocha:2024drz}. 
One case where general arguments can be made is to focus on ultrarelativistic particles at mid-rapidity, where the transverse momentum is the same as the energy, and particle multiplicity is proportional to total entropy~\cite{Gardim:2019xjs}. 
In this specific case, one can therefore argue that $\langle p_T\rangle \propto E/S$. 
If we assume that the effective volume $V_{\rm eff}$ is constant, then by definition $E/S = \varepsilon(T_{\rm eff})/s(T_{\rm eff})$, so the estimate $\langle p_T\rangle \propto E/S$ leads to a relation $\langle p_T\rangle \propto \varepsilon(T_{\rm eff})/s(T_{\rm eff})$, which yields
\begin{equation}
    \frac{{\rm d}\ln\langle p_T\rangle}{{\rm d}\ln N_{\rm ch}}  \underbrace{=}_{\substack{\langle p_T\rangle \propto E/S\\ \\ {\rm d}V_{\rm eff}=0\\ \\ N_{\rm ch} \propto S}} \frac{{\rm d}\ln(\varepsilon(T_{\rm eff})/s(T_{\rm eff}))}{{\rm d}\ln s(T_{\rm eff})}.
\end{equation}
In the absence of a chemical potential, we can write
\begin{align}
    \frac{{\rm d}\ln(\varepsilon/s)}{{\rm d}\ln s} &= \frac{{\rm d}\ln\varepsilon}{{\rm d}\ln s} - 1 = \frac{s}{\varepsilon}\frac{{\rm d}\varepsilon}{{\rm d} s} - 1 \notag \\
    &  = \frac{sT}{\varepsilon} - 1 = \frac{\varepsilon+P}{\varepsilon} - 1  = \frac{P}{\varepsilon}.
\end{align}

Hence, we have
\begin{equation}\label{eq3}
 b_{\rm UC} \equiv     \frac{{\rm d}\ln\langle p_T\rangle}{{\rm d}\ln 
 N_{\rm ch}} \underbrace{=}_{\substack{\langle p_T\rangle \propto E/S\\ \\ {\rm d}V_{\rm eff}=0\\ \\ N_{\rm ch} \propto S}}  \left.\frac{{\rm d}\ln(E/S)}{{\rm d}\ln S}\right|_{{\rm d}V_{\rm eff} = 0}  = \frac{P(T_{\rm eff})}{\varepsilon(T_{\rm eff})}
\end{equation}

This is equal to $c_s^2$ only for special equations of state when $c_s^2 = P/\varepsilon$. Indeed, if the speed of sound is constant, the energy density is $\varepsilon=A T^{1+c_s^{-2}}$, the entropy is $s=A(1{+}c_s^{2}) T^{c_s^{-2}}$ , and $\varepsilon(T_{\rm eff})/s(T_{\rm eff}) \propto T_{\rm eff}$, in which case Eq.~\eqref{eq3} reduces to Eq.~\eqref{eq:buc_cs2}  as expected.
A study of the case with a constant speed of sound can be found in Ref.~\cite{SoaresRocha:2024drz}.

From numerical simulations, $\langle p_T\rangle$ has a strong linear correlation with both $E/S=\varepsilon(T_{\rm eff})/s(T_{\rm eff})$ and with $T_{\rm eff}$, as shown in Figure~\ref{fig:pt_corr}. Hence, it is a priori not evident if one should expect to extract the speed of sound or the pressure over energy density ratio.

\subsection{The effective temperature $T_{\rm eff}$}
\label{sec:Teff_vs_mean_pT}

For Eq.~\eqref{eq:buc_cs2} or Eq.~\eqref{eq3} to provide a measurement of the thermodynamic properties of quark-gluon plasma at given temperature, one needs the precise relation between the effective temperature $T_{\rm eff}$ and $\langle p_T\rangle$.
While some general arguments have been put forward to explain why $C$ should be close to $3$~\cite{Gardim:2019xjs}, constraints on $C$ used in practice have been determined from numerical simulations:
the energy and entropy on the hypersurface obtained from a hydrodynamic simulations is used to calculate $T_{\rm eff}$ from the lattice equation of state, and is combined with $\langle p_T\rangle$ of charged hadrons from the same simulation to evaluate $C=\langle p_T\rangle/T_{\rm eff}$.
As we show in Section~\ref{sec:results}, if we use such an approach, we find $C=\langle p_T\rangle/T_{\rm eff}$ ranging from $2.7$ to $3.9$, with a significant amount of variation originating from the temperature at which the hypersurface is defined.
To constrain further the values of $C$, experimental measurements other than $b_{\rm UC}$ can be used.
For example, one can make sure that $C$ is extracted from a simulation that has been tuned to heavy-ion data in e.g.~a large scale Bayesian analysis; alternatively, one can use constraints on model parameters from specific measurements~\cite{Gardim:2019xjs}.
Previous works that used variations of this approach have quoted $C = 3.07$~\cite{Gardim:2019xjs} and $C=3\pm 0.05$~\cite{Gardim:2024zvi}, based on numerical simulations for Pb-Pb collisions constrained from various heavy-ion measurements (Refs.~\cite{Gardim:2024zvi,Gardim:2022yds} also contains information from p+Pb collisions). On the other hand, in Ref.~\cite{Mu:2025gtr} it was found that $C \approx 2.16$--$2.8$ for p+Pb collisions.
Ultimately, the exact range of values of $C$ will depend on modeling assumptions and on the data sets used to constraints the model before $C$ is extracted; one would generally expect to constrain $C$ more than the range we quote in the present work.

Experiments use these various coefficients $C$ to determine $T_{\rm eff}$ from the measured $\langle p_T\rangle$. We emphasize that, in this case, this coefficient $C$ contains information not only from the lattice equation of state, but also from the model of heavy-ion collisions used and the measurements used to constrain it.

\subsection{The effective volume $V_{\rm eff}$}
The condition that the effective volume is constant, used in every argument, is highly non--trivial.
The effective volume is not the same as the physical volume of the system. Instead, it is defined through Eq.~\eqref{eq:E_epsilon_Veff}.
One can write an expression for $V_{\rm eff}$ only in terms of $E$ and $S$:
\begin{equation}
    V_{\rm eff} (E,S) = \frac{E}{\varepsilon(s=S/V_{\rm eff})}
\end{equation}
where the function $\varepsilon(s)$ is defined by the underlying equation of state. This means that there is no way to have general knowledge about the effective volume without having prior information about the equation of state. As shown in Ref.~\cite{Sun:2024zsy}, deviations from the constant effective volume will introduce corrections to Eqs.~\eqref{eq:buc_cs2} and \eqref{eq3}. These corrections involve derivatives ${\rm d} \ln V_{\rm eff}/{\rm d}\ln S$ and thus necessarily include information from the lattice equation of state, particularly about ${\rm d} \ln \varepsilon/{\rm d} \ln s$.

\section{Model description}
\label{sec:model}
In order to test the validity of Eqs.~\eqref{eq:buc_cs2} and ~\eqref{eq3} under various assumptions, we perform boost-invariant hydrodynamic simulations using the lattice QCD equation of state by HotQCD Collaboration~\cite{Bazavov:2009zn}. As an initial condition for the temperature at proper time $\tau = 0.4$~fm, we use a Gaussian profile in the transverse-coordinate $(x, y)$ plane:
\begin{equation}
\label{eq:T-profile}
   T(\tau_0, \vec{x}_\perp) = T_0 \exp\left(-\frac{\vec{x}_\perp^2}{2\sigma^2}\right),
\end{equation}
where $\tau_0$ is the initial time and $\vec{x}_\perp$ is the position in the transverse plane. The normalization of the temperature profile is set with $T_0$, and $\sigma$ controls the transverse width of the profile. The initial transverse velocity is set to zero. We vary $T_0$ across the range $[0.23, 0.67]$~GeV in increments of $0.004$~GeV, while $\sigma$ is varied over $[4.25, 5.06]$~fm with a step size of $0.028$~fm. For each $(T_0, \sigma)$ combination, we solve the ideal hydrodynamic evolution and find constant-temperature hypersurfaces~\cite{Huovinen:2012is} for various temperatures $T_f$ using the MUSIC hydrodynamic code~\cite{Schenke:2010nt,Ryu:2015vwa,Paquet:2015lta,Ryu:2017qzn} . The final-state particles are obtained by sampling particles from these hypersurfaces using the Cooper-Frye procedure \cite{Cooper:1974mv,Cooper:1974qi} and performing decays for unstable hadrons. These are implemented using the iSS particle sampler~\cite{Shen:2014vra}. For each hypersurface, particles are repeatedly sampled until we obtain a total of 500,000 particles at midrapidity.\footnote{This is done because the sampling algorithm assumes a grand canonical ensemble, so the energy is not necessarily conserved in each individual sampling, but only on average.}

At each $(T_0, \sigma, T_f)$ point, we compute the total energy $E$ and entropy $S$ at the hypersurface $\Sigma$ as
\begin{equation}
\begin{split}
        E &= \int_\Sigma {\rm d}^3\sigma_\mu T^{t\mu},\\
        S &= \int_\Sigma {\rm d}^3\sigma_\mu s u^\mu,
\end{split}
\end{equation}
where ${\rm d}^3 \sigma^\mu$ is the directed surface element of hypersurface $\Sigma$,  $T^{\mu\nu} = (\varepsilon+p) u^\mu u^\nu - Pg^{\mu\nu}$ is the energy-momentum tensor, and $u^\mu$ is the fluid 4-velocity. The effective temperature $T_{\rm eff}$ and the effective volume $V_{\rm eff}$ can then be computed from $E$ and $S$ using Eq.~\eqref{eq:E_epsilon_Veff}. We once again remark that, in general, solving for $T_{\rm eff}$ and $V_{\rm eff}$ requires knowing the equation of state.

Now that we know $E, S, \langle p_T\rangle, N,$ and $V_{\rm eff}$  at every $(T_0, \sigma, T_f)$ point, we construct 2D-spline interpolators in the $(T_0, \sigma)$ plane for each of these at each hypersurface of constant temperature $T_f$. These interpolators can be used to differentiate observables with respect to $T_0$ and $\sigma$. To compute the variations in observables, one needs to choose how these variations arise from the initial state. In practice, for each $T_f$, all of our variations depend only on the initial temperature $T_0$ and the width $\sigma$ of the initial transverse temperature profile, so for any observable $\mathcal{O}$, we can write:
\begin{equation}
   \mathrm{d}\mathcal{O} = \frac{\partial \mathcal{O}}{\partial T_0} \mathrm{d} T_0 + \frac{\partial \mathcal{O}}{\partial \sigma} \mathrm{d} \sigma,
\end{equation}
and therefore we only need to choose variations $\mathrm{d} T_0$ and $\mathrm{d} \sigma$. Note that when computing quantities like ${\rm d}\ln(E/S)/{\rm d}\ln S$ only the ratio $d\sigma/dT_0$ matters. In the following, we use two different choices:
\begin{itemize}
    \item The first option is to choose variations of $T_0$ and $\sigma$ such that $V_{\rm eff}$ remains constant, i.e., ${\rm d}V_{\rm eff} = 0$, which is consistent with the assumption made in Eqs.~\eqref{eq:buc_cs2} and ~\eqref{eq3}.
    \item  The second option is to fix $\mathrm{d}\sigma = 0$ and vary only $T_0$. This approach is similar to one taken in Refs.~\cite{SoaresRocha:2024drz, Gardim:2024zvi}, but it does not guarantee that ${\rm d}V_{\rm eff} = 0$. In fact, $V_\text{eff}=V_\text{eff}(E,S)$, where $E$ and $S$ depend on $T_0$ in a non-trivial way, so that in general $(\partial V_\text{eff}/\partial T_0)_{\sigma}\neq 0$.  
\end{itemize}
We note that choosing the variation along which one computes the derivatives is not unique to our model, but it also appears in more realistic event-by-event simulations. There, one needs to decide how to order the events in centrality classes before averaging over the events, and different choices can lead to different results~\cite{Nijs:2023bzv}.

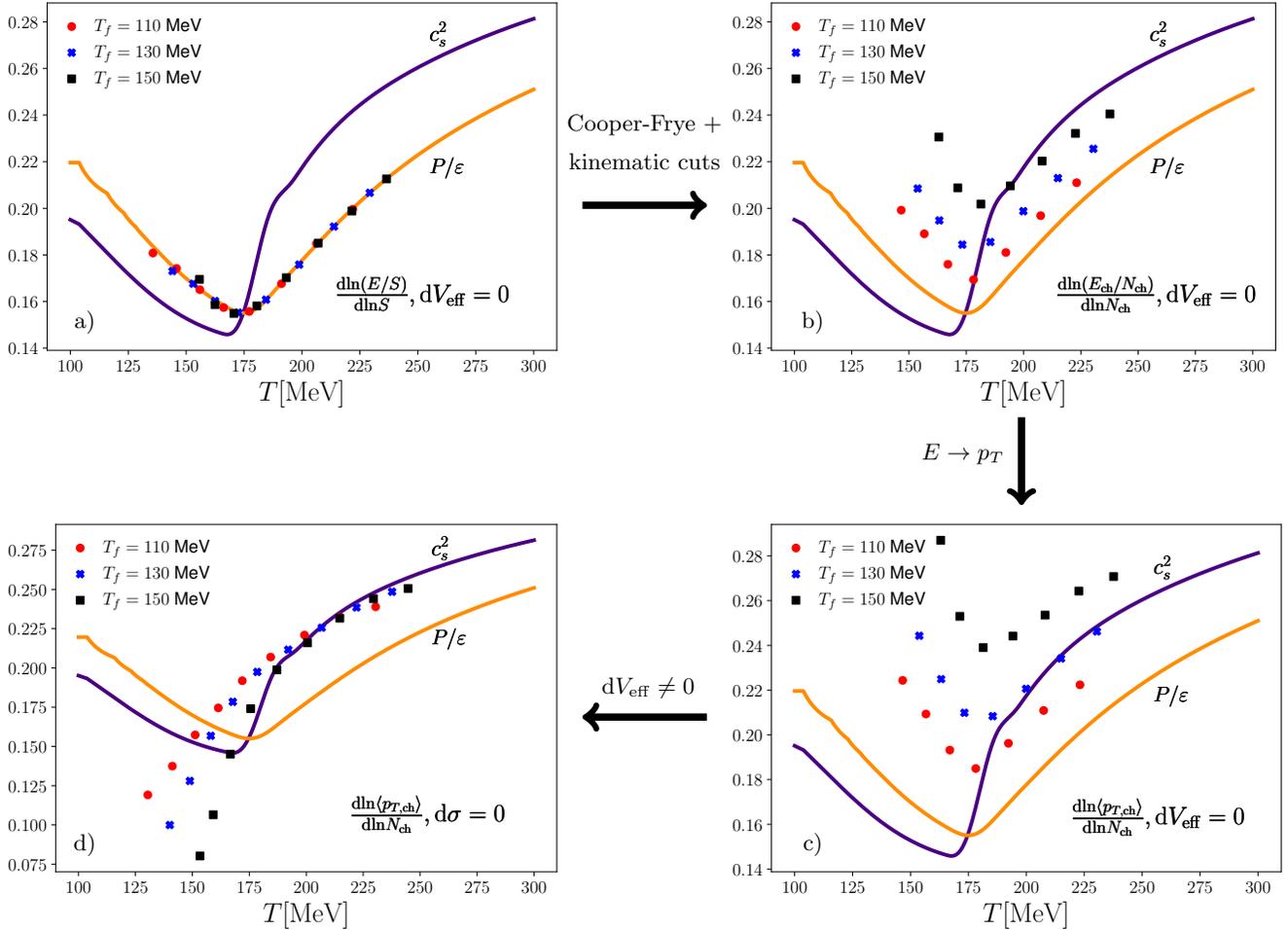
\begin{figure*}
    \begin{tikzpicture}
    \node (a) at (0.0, 0.0) {\includegraphics[width=.91\columnwidth]{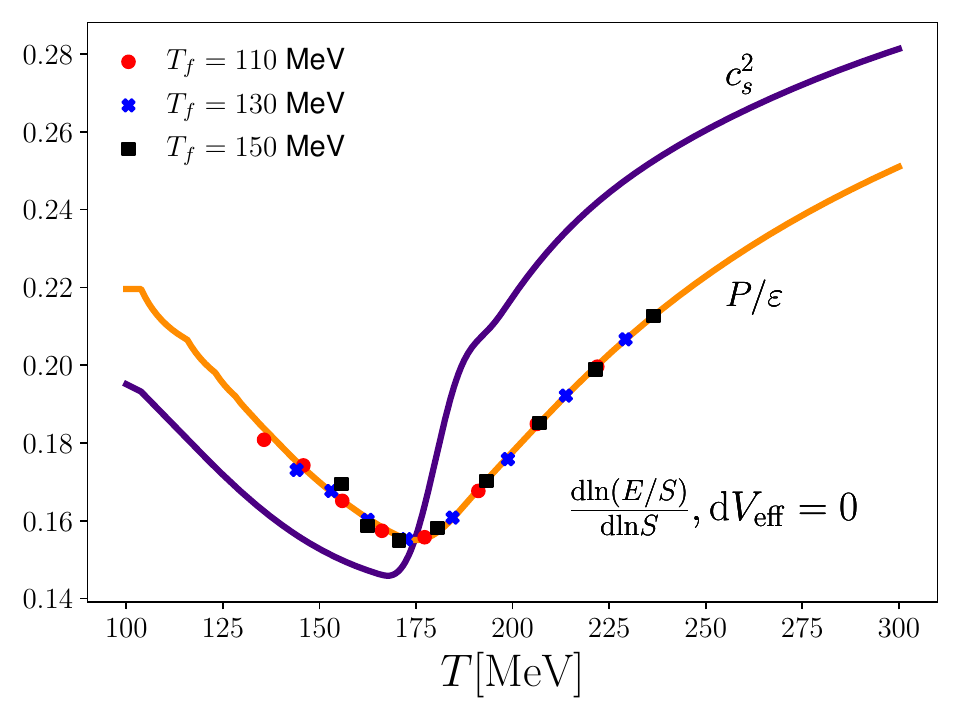}};
    \node (b) [below=of a] {\includegraphics[width=.91\columnwidth]{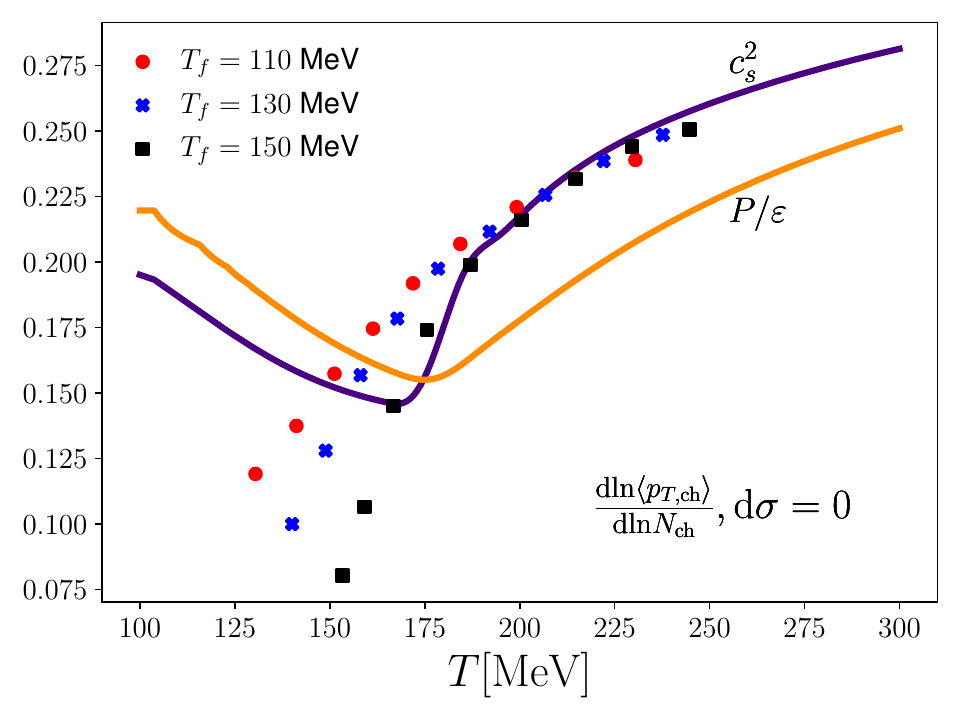}};
    \node (c) [right=of b] {\hspace{8mm}\includegraphics[width=.91\columnwidth]{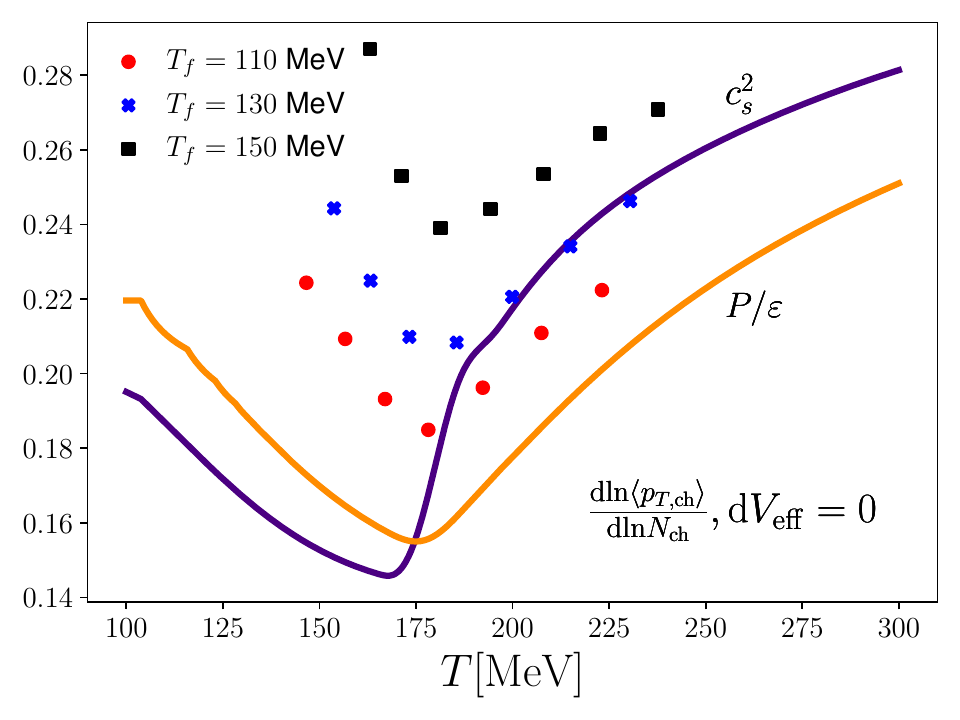}};
    \node (d) [right=of a] {\hspace{8mm}\includegraphics[width=.91\columnwidth]{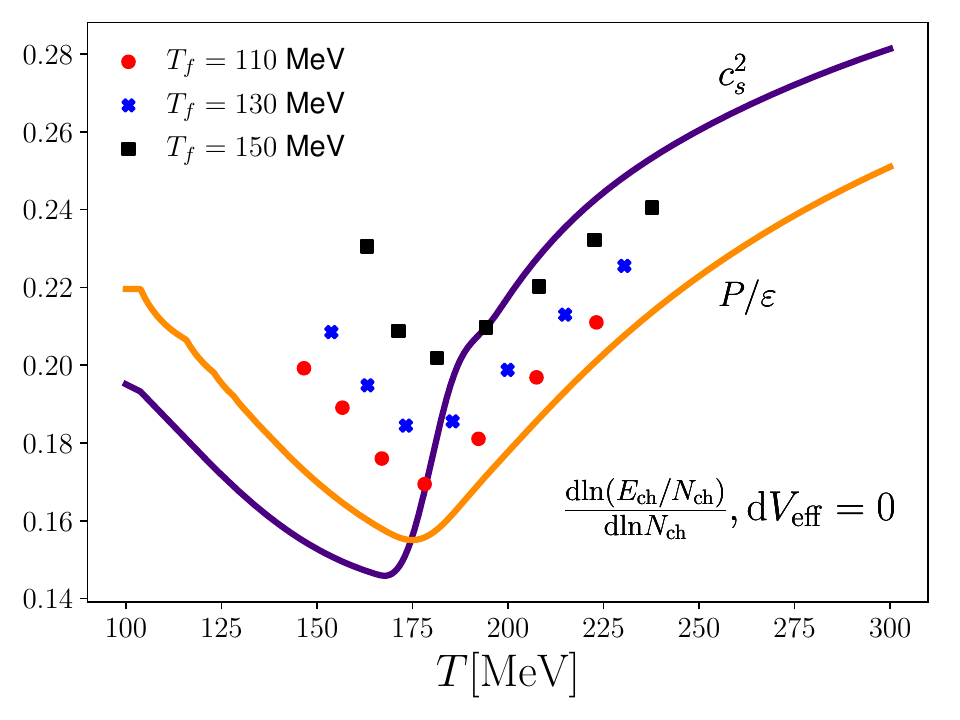}};
    \draw [line width=1mm ,black,->] (10.1,-2.9) to  (10.1, -4.1); 
    \draw [line width=1mm,black,->] (4.1, 0.0) to  (5.8, 0.0);
    \draw [line width=1mm,black,->] (5.8, -7.0) to  (4.1, -7.0);
    \node (e) at (4.95, 1.1) {Cooper-Frye +};
    \node at (4.95, 0.6) {kinematic cuts };
    \node (f) at (9.3, -3.4) {$E \to p_T$};
    \node (g) at (5.0, -6.6) {${\rm d}V_{\rm eff} \neq 0$};
    \node at (-2.7, -1.6) {a)};
    \node at (7.25, -1.6) {b)};
    \node at (7.25, -8.75) {c)};
    \node at (-2.7, -8.75) {d)};    
    \end{tikzpicture}    
\caption{Accumulation of effects that appears to coincidentally lead from $P/\varepsilon$ to $c_s^2$ for temperatures between 200 and 250 MeV. Panel a) Assuming $V_\text{eff}$ constant, the derivative $d\ln(E/S)/d\ln S$ computed from the energy $E$ and entropy $S$ on constant-temperature hypersurfaces leads to $P/\varepsilon$, and not $c_s^2$. Panel b) When we replace the hydrodynamic energy on the hypersurface with the Cooper-Frye particle energy with midrapidity cuts, a first error is generated, which scatters the point above $P/\varepsilon$. Panel c) When we replace the mean energy $E/S$ with $\langle p_T \rangle$, the points are scattered even more randomly above $P/\varepsilon$. Panel d) When we finally relax the assumption that $V_\text{eff}=\text{const}$ (which was a central ingredient in the original derivation by \cite{Gardim:2019xjs,Gardim:2019brr}), and fix instead the width $\sigma$ of the initial profile, the points rearrange on a curve that happens to fall on top of the $c_s^2$-curve between temperatures 200 and 250 MeV.}
    \label{fig}
\end{figure*}

\section{Results}
\label{sec:results}
\subsection{QCD equation of state}
We begin by evaluating ${\rm d}\ln(E/S)/{\rm d}\ln S$ on hypersurfaces with three different temperatures $T_{f}$ and various values of $T_0$. Here we choose a reference value of $\sigma =$ 4.53 fm, but we have checked that changing $\sigma$ causes only minor changes. For each point, we compute $T_{\rm eff}$ and choose variations $dT_0$ and  $d\sigma$ such that ${\rm d}V_{\rm eff} = 0$. In Fig.~\ref{fig}a), results obtained this way are compared against the speed of sound squared $c_s^2$ and $P/\varepsilon$ from the HotQCD lattice equation of state~\cite{Bazavov:2009zn}.  Since ${\rm d}V_{\rm eff} = 0$, the last equality in Eq.~\eqref{eq3} holds and we recover $P/\varepsilon$ exactly, as expected.

In Fig. \ref{fig}b), we show ${\rm d}\ln\langle (E_{\rm ch}/N_{\rm ch})\rangle/{\rm d}\ln N_{\rm ch}$ evaluated from charged hadrons at midrapidity with pseudorapidity $\abs{\eta}\leq0.5$. In going from Fig. \ref{fig}a) to \ref{fig}b), we have implemented the Cooper-Frye procedure, and the appropriate kinematic cuts on the pseudorapidity of particles. This is an essential distinction from the previous step, since even for a boost invariant system at mid-rapidity, the total energy content of particles in a unit of pseudorapidity is not the same as the total energy on the hypersurface in a unit of spacetime rapidity. While the effective temperature $T_{\rm eff}$ is defined using the energy and entropy calculated using the spacetime-rapidity window $\eta_{s} \in [-0.5, 0.5]$, the observable ${\rm d}\ln\langle (E_{\rm ch}/N_{\rm ch})\rangle/{\rm d}\ln N_{\rm ch}$ is calculated using particles with kinematic cuts on momentum pseudorapidity. 
Due to this reason, the results no longer agree with $P/\varepsilon$, but instead we get a $T_f$ dependent shift above the $P/\varepsilon$ curve. Importantly, the minima of ${\rm d}\ln\langle (E_{\rm ch}/N_{\rm ch})\rangle/{\rm d}\ln N_{\rm ch}$ for the different constant-temperature hypersurfaces are observed to be closer to the minimum of $P/\varepsilon$ than the minimum of $c_s^2$.

We point out that in the late time limit, when $\tau\rightarrow\infty$ and $T_f \rightarrow0$, the difference between the energy content on the hypersurface in a unit of spacetime-rapidity and the total energy of particles in a unit of pseudorapidity vanishes. As the hypersurface temperature is decreased, the high-momentum part of the spectrum is suppressed in the local rest frame of the fluid. In the limit $T_f\rightarrow0$, all the momentum of the particles come from the flow of the underlying medium. In that limit, we should recover $P/\varepsilon$ from the observable $b_{\rm UC}$. See \cite{SoaresRocha:2024drz} for a discussion on the late time limit of hydrodynamic evolution.\footnote{We note that in Ref.~\cite{SoaresRocha:2024drz} $P/\varepsilon = c_{s}^{2}$} The same pattern is also seen in Fig.~\ref{fig}b), where lowering the hypersurface temperature leads to a better agreement with $P/\varepsilon$.

Figure \ref{fig}c) shows the observable $b_{\rm UC}$ computed from the charged hadron average transverse momentum and multiplicity  --- that is, in going from Fig. \ref{fig}b) to c), we moved from using the mean energy of the particles to their mean transverse momentum. The latter is more easily accessible in experiments. This brings in additional corrections from the rest mass of particles in Eq. \eqref{eq3}. We again see that there is a hypersurface temperature $T_f$ dependence in $b_{\rm UC}$ and it has shifted further above the $P/\varepsilon$ curve. However, the temperature dependence of the $b_{\rm UC}$ still closely resembles the shape of $P/\varepsilon$. 
Moreover, Fig. \ref{fig}c) proves that in general, the statements $\langle p_T \rangle \propto T_{\rm eff}$ and $N_{\rm ch} \propto S$ are not both sufficiently accurate, since otherwise all points would agree with $c_s^2$ as per Eq.~\eqref{eq:buc_cs2}.  If ${\rm d}V_{\rm eff} = 0$ was actually true across different ultracentral events, Fig. \ref{fig}c) is the observable that we would expect to measure in experiments: a quantity correlated with $P/\varepsilon$ and $c_s^2$, but not an accurate measurement of either. 

Finally, in going from Fig.~\ref{fig}~c) to d) we relax the requirement ${\rm d}V_{\rm eff}=0$. This effectively means that we just vary the initial normalization $T_0$ in our simulations and keep the Gaussian width $\sigma$ of the initial temperature profile fixed. 
As seen in Fig. \ref{fig}d), the dependence of the observable on the temperature $T_f$ of the hypersurface is much smaller. 
This appears to be an accidental cancellation of effects from kinematic cuts and $V_{\rm eff}$ dependence. The observable $b_{\rm UC}$ is again shifted above the $P/\varepsilon$ curve and is close to $c_s^2$ for temperatures between 180 MeV to 230 MeV. At lower temperatures, there is a noticeable dependence on the hypersurface temperature and the observable is inconsistent with $c_s^2$. 

The agreement between $b_{\rm UC}$ and $c_s^2$ in the narrow temperature region appears to be a coincidental accumulation of effects from assumptions $\langle p_T\rangle\propto E/S=\varepsilon(T_{\rm eff})/s(T_{\rm eff})$ and ${\rm d}V_{\rm eff} = 0$ and we have no reason to believe that this is a general feature. In fact, the case with ${\rm d}V_{\rm eff} = 0$ is noticeably farther from $c_s^2$.

\subsection{Modified equation of state}

\begin{figure*}
    \begin{tikzpicture}
    \node (a) at (0.0, 0.0) {\includegraphics[width=.91\columnwidth]{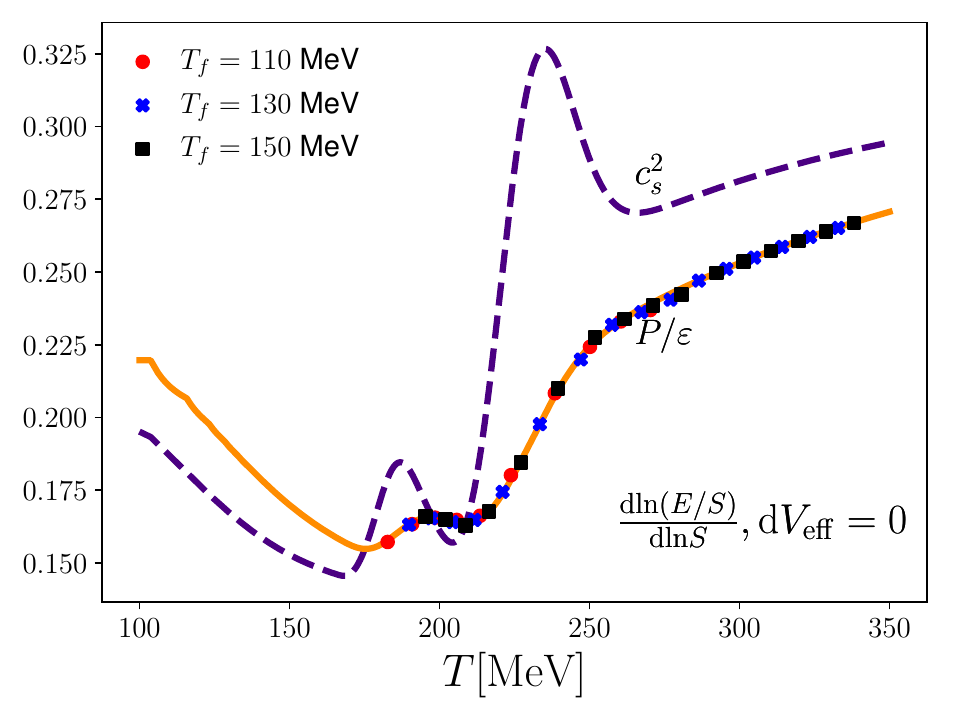}};
    \node (b) [below=of a] {\includegraphics[width=.91\columnwidth]{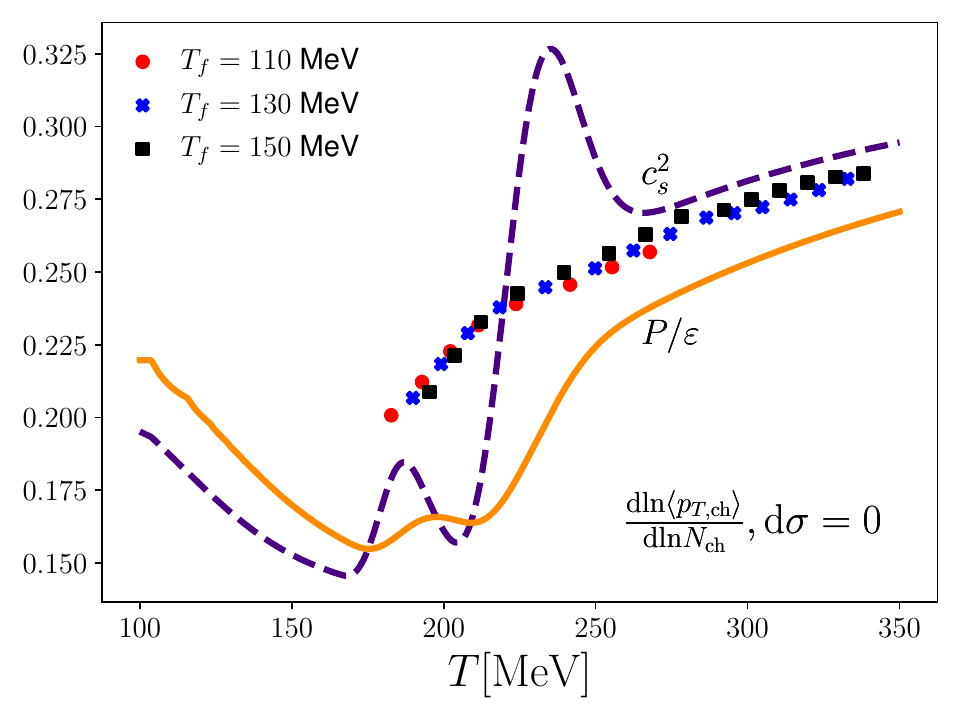}};
    \node (c) [right=of b] {\hspace{8mm}\includegraphics[width=.91\columnwidth]{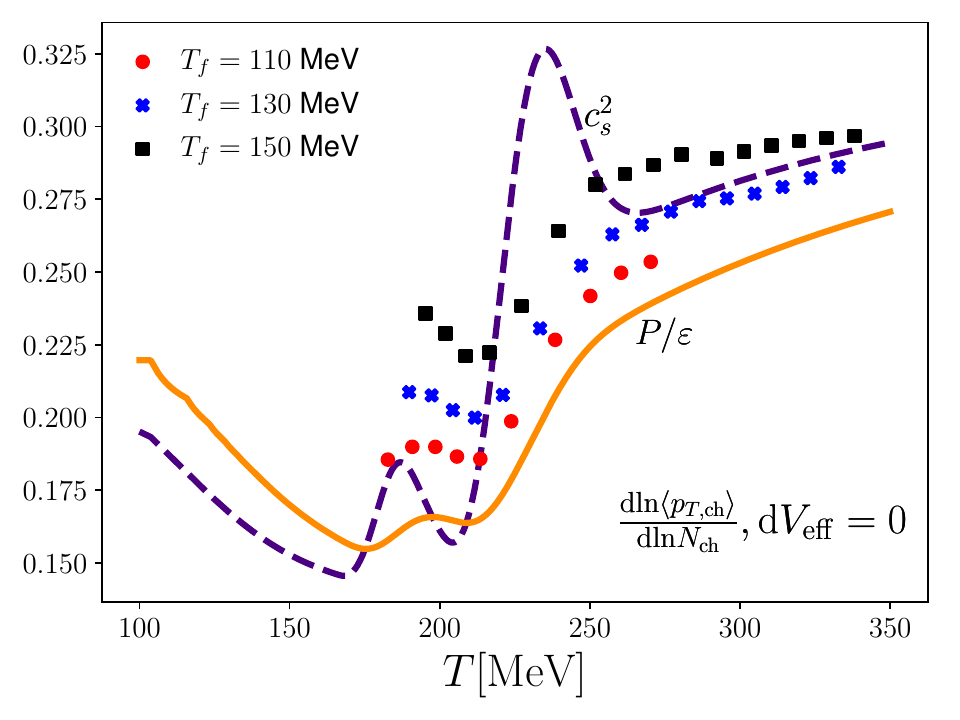}};
    \node (d) [right=of a] {\hspace{8mm}\includegraphics[width=.91\columnwidth]{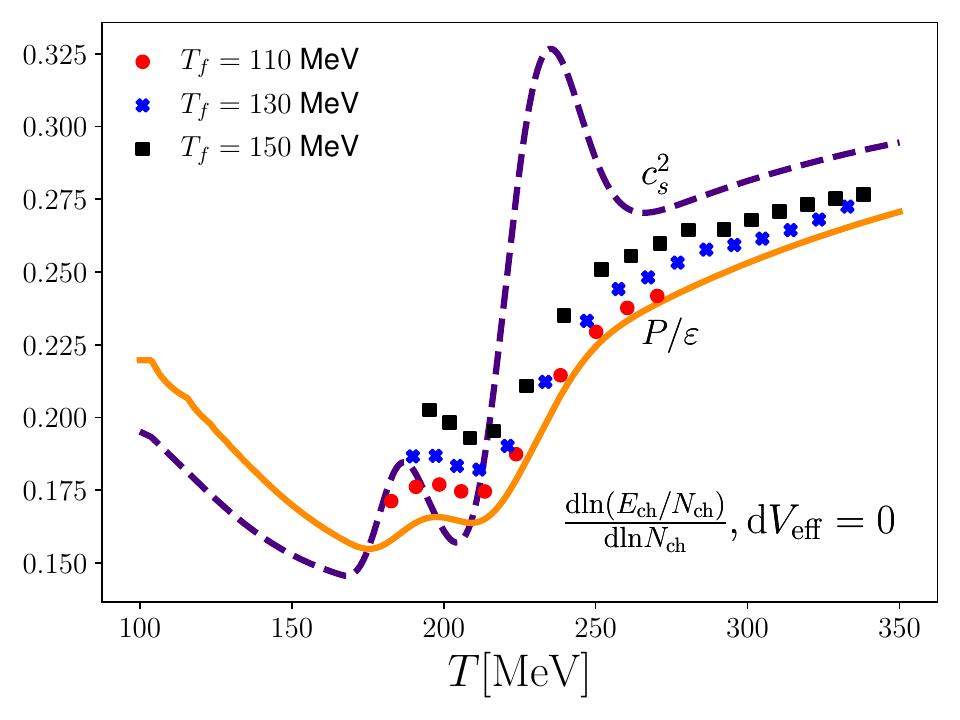}};
    \draw [line width=1mm ,black,->] (10.1,-2.9) to  (10.1, -4.1); 
    \draw [line width=1mm,black,->] (4.1, 0.0) to  (5.8, 0.0);
    \draw [line width=1mm,black,->] (5.8, -7.0) to  (4.1, -7.0);
    \node (e) at (4.95, 1.1) {Cooper-Frye +};
    \node at (4.95, 0.6) {kinematic cuts };
    \node (f) at (9.3, -3.4) {$E \to p_T$};
    \node (g) at (5.0, -6.6) {${\rm d}V_{\rm eff} \neq 0$};
    \node at (-2.7, -1.6) {a)};
    \node at (7.25, -1.6) {b)};
    \node at (7.25, -8.75) {c)};
    \node at (-2.7, -8.75) {d)};    
    \end{tikzpicture}    
\caption{Same as Fig~\ref{fig} but with a modified equation of state for temperatures $T>180$ MeV, see Eq.~\eqref{eq:cs2-modif}. The systematics for panels a), b) and c) are comparable to those seen earlier, while observable $b_{\rm UC}$ in panel d) shows much larger deviations from the speed of sound $c_s^2$.  }
    \label{fig2}
\end{figure*}

To further test the agreement between the observable $b_{\rm UC}$ and $c_s^2$, we repeated the same procedure as earlier, but this time modifying the equation of state. For our modified equation of state, the hadronic part with $T \leq 150$ MeV is left unmodified, while for $T > 150$ MeV, the speed of sound is given by
\begin{equation}
\label{eq:cs2-modif}
    c_s^2(T) = c_{s,\rm QCD}^2 (T)\left[1+35\tanh\left(\frac{T-T_{a}}{\rm GeV} \right) e^{-\frac{\left(T-T_{a} \right)^2}{\delta T^2}} \right],
\end{equation}
where $T_a = 220$ MeV, and $\delta T = 20$ MeV and $c_{s,\rm QCD}^2$ is the speed of sound from the QCD equation of state used earlier. Other thermodynamic quantities above temperature 150 MeV are then constructed from $c_s^2(T)$ in such a way that all thermodynamic identities are satisfied. While using the modified equation of state, the initial state parameters were varied between values $[0.45, 0.96]$~GeV for $T_0$, and $[4.38, 5.06]$ fm for $\sigma$.

In Fig.~\ref{fig2}, we show the same set of plots as discussed in the previous subsection, but this time with our modified equation of state. When ${\rm d}V_{\rm eff}=0$  our previous observations hold. Firstly, for ${\rm d}\ln(E/S)/{\rm d}\ln S$ in Fig.~\ref{fig2}a), the points agree with $P/\varepsilon$, regardless of the hypersurface temperature $T_{f}$. Secondly, when computing ${\rm d}\ln\langle (E_{\rm ch}/N_{\rm ch})\rangle/{\rm d}\ln N_{\rm ch}$ (Fig.~\ref{fig2}b)), each set of points for a given $T_{f}$ display a pattern similar to $P/\varepsilon$, with lower $T_{f}$ having better quantitative agreement. Thirdly, the mass corrections due to using transverse momentum instead of energy lead to slight behavior changes in Fig.~\ref{fig2}c) with respect to Fig.~\ref{fig2}b). 
However, in Fig.~\ref{fig2}d) when computing the exponent $b_{\rm UC}$ by only varying the normalization of initial conditions (changing $T_0$ but keeping $\sigma$ fixed in Eq.~\eqref{eq:T-profile}), we no longer see a correlation between  $b_{\rm UC}$ and $c_s^2$ in the 180-230 MeV temperature range. 
For the very high temperatures $b_{\rm UC}$ again approaches $c_s^2$.
From Figs.~\ref{fig} and \ref{fig2}, it appears that $b_{\rm UC}$ tracks $c_s^2$ more closely at high effective temperatures where $c_s^2$ is varying slowly. It is not clear if this is related to these high $T_{\rm eff}$ regions being closer to the conformal ($c_s^2=1/3$) limit, or to high $T_{\rm eff}$ systems being longer lived, two effects studied in Ref.~\cite{SoaresRocha:2024drz}. The significant effect of enforcing a constant effective volume --- panels (c) and (d) --- seems to preclude a simple explanation.

\subsection{The relation between $\langle p_T\rangle$ and $T_{\rm eff}$}

\begin{figure*}
    \begin{tikzpicture}
    \centering
    \node (a) at (0.0,0.0) {\includegraphics[width=0.45\linewidth]{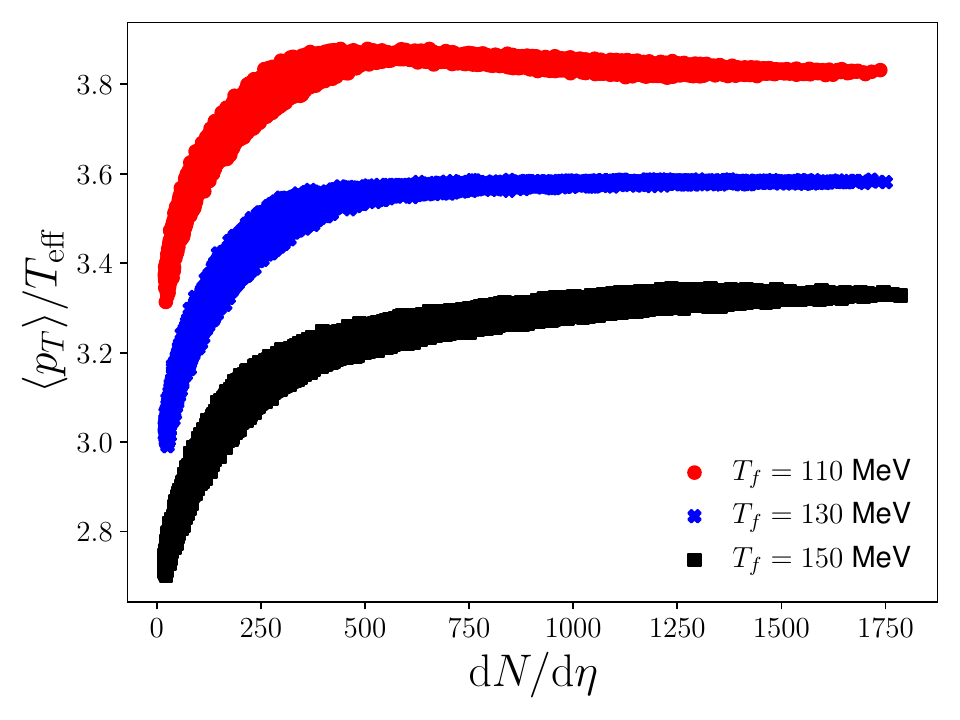}};
    \node (b) [right=of a] {\includegraphics[width=0.45\linewidth]{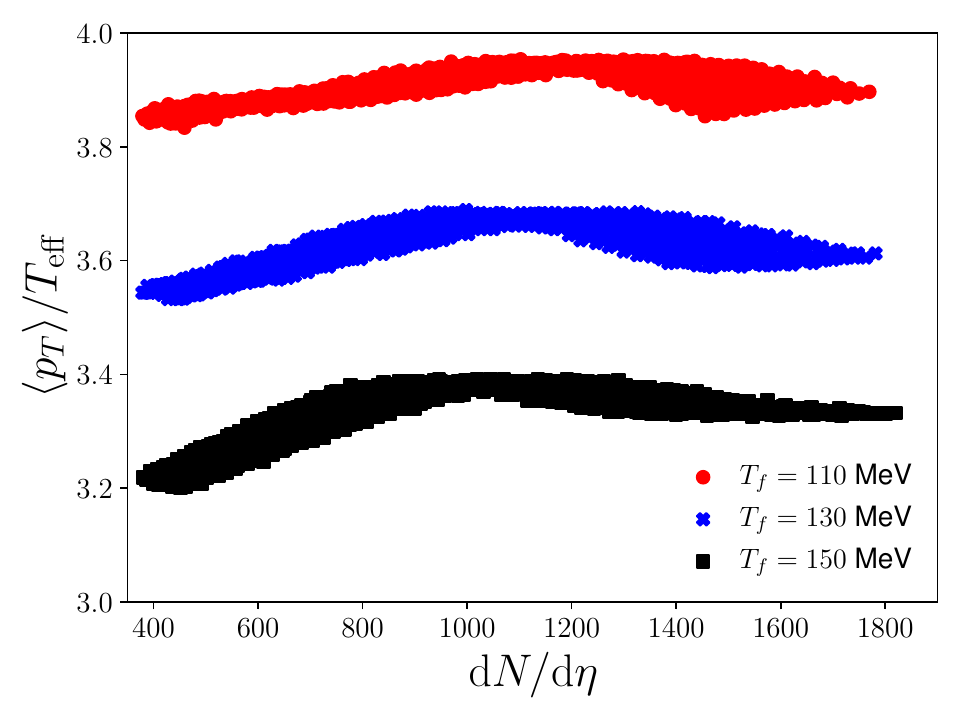}};
    \node at (-1.5, - 1.5) {\text{QCD EoS}};
    \node at (8.0, - 1.5) {\text{Modified EoS}};
    \end{tikzpicture}
    \caption{The $\langle p_T \rangle /T_{\rm eff}$ ratio as a function of charged hadron multiplicity for QCD equation of state (left-hand panel) and the modified equation of state (right-hand panel).}
    \label{fig3}
\end{figure*}

We also studied the validity of assumption $\langle p_T\rangle \propto T_{\rm eff}$. In Fig.~\ref{fig3}, we show $\langle p_T \rangle/T_{\rm eff}$ as a function of charged particle multiplicity for the QCD equation of state (left panel) and the modified equation of state (right panel, see Eq.~\eqref{eq:cs2-modif}). We notice that the ratio $\langle p_T \rangle/T_{\rm eff}$ has a clear dependence on the temperature of the hypersurface in both cases. Besides, in the left panel of Fig.~\ref{fig3}, we see that for sufficiently large multiplicities, the range of values of $\langle p_T \rangle/T_{\rm eff}$ lies within a horizontal band whose width depends non-trivially on the hypersurface temperature. We see that these bands lie around $\langle p_T \rangle/T_{\rm eff} \approx 3.2$ for $T_{f} = 150$ MeV, $\langle p_T \rangle/T_{\rm eff} \approx 3.5$ for $T_{f} = 130$ MeV, and $\langle p_T \rangle/T_{\rm eff} \approx 3.8$ for $T_{f} = 110$ MeV. The variation in the values of $C$ obtained is strong evidence that the coefficient cannot be constrained precisely from general theoretical arguments nor from simulations alone: some form of constraints from data (others than $b_{U C}$) is required (see Section~\ref{sec:Teff_vs_mean_pT}).
For ${\rm d} N/{\rm d} \eta \lesssim 500$, the band is such that $\langle p_T \rangle/T_{\rm eff}$ is not approximately constant, but rather decreases as the multiplicity decreases. The effect of the hypersurface temperature and multiplicity dependence on $\langle p_T \rangle /T_{\rm eff}$ has also been seen in Refs.~\cite{Gardim:2019xjs,Gardim:2024zvi}.

With the modified equation of state (Eq.~\eqref{eq:cs2-modif}), in the right panel of Fig.~\ref{fig3}, the shape of the bands changes significantly: they become wider at high multiplicities and possess a different range in multiplicity even though the range of values of $T_{0}$ and $\sigma$ are very similar (see, respectively, the text below Eq.~\eqref{eq:T-profile} and the text below Eq.~\eqref{eq:cs2-modif}). 


We emphasize that even in the situations where $\langle p_T \rangle/T_{\rm eff}$ ratio seems to be almost a constant, it does not guarantee that the observable $b_{\rm UC}$ would be $c_s^2$ even when ${\rm d}V_{\rm eff}=0$ (see Fig.~\ref{fig}c)). The reason for this is the following. 
It is possible to write 
\begin{align}
b_{\rm UC} = {\rm d ln}~T_{\rm eff}/{\rm d ln} N + {\rm d ln}(\langle p_T \rangle /T_{\rm eff})/{\rm d ln} N,     
\end{align}
where the first term is $c_s^2$, if $N \propto S$ and ${\rm d} V_{\rm eff} = 0$. When the $\langle p_T \rangle/T_{\rm eff}$ ratio is exactly constant, the second term would be zero and $c_s^2$ is recovered. However, since the $\langle p_T \rangle/T_{\rm eff}$ forms a finite band even at high multiplicities, its derivative can get slightly different values depending on what is the underlying variation that causes the changes in the multiplicity and mean transverse momentum. It turns out that when we choose a variation for which ${\rm  d}V_{\rm eff} = 0$, then ${\rm d ln}(\langle p_T \rangle /T_{\rm eff})/{\rm d ln} N$ term modifies the functional form of $b_{\rm UC}$ significantly.

\section{Discussion} 
\label{sec:discussion}
Heavy-ion collisions are evidently not simple thermodynamic systems, and the hypersurface energy $E$ and entropy $S$ are not experimentally measurable. What is measured is typically the mean transverse momentum $\langle p_{T,{\rm ch}}\rangle$ and the multiplicity of charged hadrons $N_{\rm ch}$ in a specific range of pseudorapidity. In ultracentral collisions, these are found to be related by $\langle p_{T,{\rm ch}}\rangle \propto N_{\rm ch}^{b_{\rm UC}}$ with $b_{\rm UC}$ a number in the vicinity of $1/4$.
The observable $b_{\rm UC}$ has been previously assumed to be so closely related to the speed of sound as to provide constraints on the equation of state of QCD that are competitive with lattice QCD~\cite{CMS:2024sgx}.
As discussed in this work, this interpretation of $b_{\rm UC}$ is not supported by closer scrutiny. Under specific conditions ($N \propto S$, $\langle p_{T} \rangle \propto E/S$, no rapidity cut, and ${\rm d}V_{\rm eff}=0$), the exponent $b_{\rm UC}$ should be measuring the ratio of the pressure and the energy density $P/\varepsilon$ at the effective temperature $T_{\rm eff}$ defined by Eq.\eqref{eq:E_epsilon_Veff}. As seen in Figure~\ref{fig}, we found evidence that this is not possible in reality.

On the other hand, under other conditions, $b_{\rm UC}$ does appear to match the speed of sound --- the original interpretation for $b_{\rm UC}$. Specifically, if the condition ${\rm d}V_{\rm eff} = 0$ is relaxed such that the variations of $\langle p_T\rangle$ and $N$ arise purely from the normalization of initial temperature $T_0$, the result from numerical simulations for $b_{\rm UC}$ is \emph{sometimes} rather close to the speed of sound (Figs.~\ref{fig}(d) and~\ref{fig2}(d)). 
This is consistent with results from other groups' numerical simulations~\cite{Gardim:2019xjs,Gardim:2024zvi}, where ${\rm d}V_{\rm eff} = 0$ was also \emph{not} enforced.

However, our results indicate that the agreement of $b_{\rm UC}$ with the speed of sound in Figs.~\ref{fig}(d) and~\ref{fig2}(d) is not a general result, for multiple reasons:
\begin{itemize}
    \item $b_{\rm UC}$ and $c_s^2$ only agree in certain ranges of temperature, and large deviations appear at low temperature (Figure~\ref{fig}(d)),
    \item $b_{\rm UC}$ and $c_s^2$ only agree if the condition ${\rm d}V_{\rm eff} = 0$ is relaxed (Figure~\ref{fig}(d) vs Figure~\ref{fig}(c)), while theoretical arguments for this relation rely on the condition ${\rm d}V_{\rm eff} = 0$,
    \item If the equation of state is modified, much larger deviations between $b_{\rm UC}$ and $c_s^2$ can be observed, as shown in Figure~\ref{fig2}.
\end{itemize}

We believe that these three observations confirm that $c_s^2(T_{\rm eff})$ cannot be equated to $b_{\rm UC}$ in general. In fact, it appears that the only way to know if $b_{\rm UC}$ accurately tracks the speed of sound is to have verified that relation ahead of time with numerical simulations that use the lattice equation of state.
We emphasize that the tests above ignore other expected challenges, such as the effect of viscosity, $p_T$ cuts, etc. discussed in other publications~\cite{Nijs:2023bzv,Mu:2025gtr}.

Additional insights from numerical simulations can help find less direct but more accurate relations between $b_{\rm UC}$ and the equation of state, as explored in Refs.~\cite{Sun:2024zsy,Mu:2025gtr} for example. 
It is our understanding that $b_{\rm UC}$ was never an independent measurement of the speed of sound, since (i) the relation between $\langle p_T \rangle$ and $T_{\rm eff}$ was evaluated using simulations that employed the lattice QCD equation of state and parameters that were tuned to heavy-ion data, and (ii) it does not appear possible to know ahead of time if $b_{\rm UC}$ and the speed of sound will agree without numerical simulations that use the lattice equation of state.
Hence, we believe that adding more information from numerical simulations (e.g.~volume variation corrections) to obtain a more complex relation between $b_{\rm UC}$ and QCD thermodynamic quantities is consistent with the original approach. Direct comparison of numerical simulations to multiplicity and average transverse momentum data from ultracentral collisions is evidently also an effective approach to leverage the thermodynamic information available in these collisions.

In summary, from numerical simulations and experimental data, it appears to be generally true  that $\langle p_{T,{\rm ch}}\rangle \propto N_{\rm ch}^{b_{\rm UC}}$ and that the values of ${b_{\rm UC}}$ extracted --- broadly between 0.15 and 0.3 --- are in the range  expected for the speed of sound of QCD or the ratio $P/\varepsilon$ of the pressure over the energy density. The prediction~\cite{Gardim:2019xjs} that the exponent $b_{\rm UC}$ should be numerically similar to the speed of sound of QCD is clearly a success of our understanding of relativistic heavy-ion collisions. However, turning this argument around and promoting $b_{\rm UC}$ as a systematic and accurate measurement of the speed of sound is a completely different matter that is not supported by our study.

\begin{center}
    \textbf{Acknowledgments}
\end{center}

The authors thank Fernando Gardim, Andre V.~Giannini, Wei Li, Andi Mankolli, Jean-Yves Ollitrault, Shengquan Tuo, Julia Velkovska and Li Yan for useful discussions. H.H., J.-F.~P., M.~S. and G.~S.~R. are supported by Vanderbilt University and by the U.S. Department of Energy, Office of Science under Award Number DE-SC-0024347.
H.H. is also partly supported by the U.S. Department of Energy, Office of Science under Award Number DE-SC0024711, and the National Science Foundation under Grant No. DMS-2406870.
L.G. is partially supported by a Vanderbilt Seeding Success grant.

\bibliographystyle{apsrev4-1}
\bibliography{refs}

\end{document}